\documentclass[10pt,conference,epsfig]{IEEEtran}
\IEEEoverridecommandlockouts
\usepackage{cite}
\usepackage{amsmath,amssymb,amsfonts}
\usepackage{algorithmic}
\usepackage{graphicx}
\usepackage{textcomp}
\usepackage{xcolor}
\usepackage{stfloats}
\usepackage{setspace}
\usepackage{graphicx}
\usepackage{subfig}
\usepackage{booktabs} 
\usepackage{siunitx}    
\newcommand{\mv}[1]{\mbox{\boldmath{$ #1 $}}}
\usepackage{geometry}
\begin{document}

\title{ CNN-Based Channel Map Estimation for Movable Antenna Systems}
\author{
	\IEEEauthorblockN{
		Yitai Huang\IEEEauthorrefmark{1}, 
		Weidong Mei\IEEEauthorrefmark{2}, 
		Xin Wei\IEEEauthorrefmark{2}, 
		Zhi Chen\IEEEauthorrefmark{2} 
		and Boyu Ning\IEEEauthorrefmark{2}} 
        \thanks{This work was supported in part by the National Key Research and Development Program of China under Grant {2024YFB2907900}.}
	\IEEEauthorblockA{\IEEEauthorrefmark{1}School of Information and Communication Engineering,}
    \IEEEauthorblockA{\IEEEauthorrefmark{2}National Key Laboratory of Wireless Communications,\\University of Electronic Science and Technology of China, Chengdu, China.}
    Emails: elton@std.uestc.edu.cn, wmei@uestc.edu.cn, xinwei@std.uestc.edu.cn,\\ chenzhi@uestc.edu.cn, boydning@outlook.com}

\maketitle
\begin{abstract}
Movable antenna (MA) has attracted increasing attention in wireless communications due to its capability of wireless channel reconfiguration through local antenna movement within a confined region at the transmitter/receiver. However, to determine the optimal antenna positions, channel state information (CSI) within the entire region, termed small-scale channel map, is required, which poses a significant challenge due to the unaffordable overhead for exhaustive channel estimation at all positions. To tackle this challenge, in this paper, we propose a new convolutional neural network (CNN)-based estimation scheme to reconstruct the small-scale channel map within a three-dimensional (3D) movement region. Specifically, we first collect a set of CSI measurements corresponding to a subset of MA positions and different receiver locations offline to comprehensively capture the environmental features. Subsequently, we train a CNN using the collected data, which is then used to reconstruct the full channel map during real-time transmission only based on a finite number of channel measurements taken at several selected MA positions within the 3D movement region. Numerical results demonstrate that our proposed scheme can accurately reconstruct the small-scale channel map and outperforms other benchmark schemes.
\end{abstract}

\section{Introduction}
The evolution of mobile communications has progressed from the use of multiple-input multiple-output (MIMO) systems in 4G to the implementation of massive MIMO in 5G, paving the way for the potential emergence of extremely large-scale MIMO (XL-MIMO) systems in the upcoming 6G era\cite{wang2024tutorial}. However, despite ongoing advancements in antenna miniaturization, accommodating an increasing number of antennas over high frequencies remains difficult due to the high hardware cost and radio-frequency (RF) energy consumption\cite{ning2023beamforming}. 

Recent advances in movable antenna (MA) technology present a promising solution to address this limitation. By leveraging the local antenna movement within a confined region, MA systems can achieve a comparable or even better performance than conventional fixed-position antenna (FPA) systems with a much smaller number of RF chains and antennas, thus significantly reducing the hardware cost and energy consumption\cite{zhu2023movable,zhu2023modeling,cui2022channel,zhu2023movable2,mei2024movable,10579873,10901030,10901621}. However, to fully unlock the potential of MAs, it is essential to obtain the channel state information (CSI) for any pair of antenna positions within the transmit and receive regions, referred to as {\it{small-scale channel map}}. This poses a significant challenge compared to channel estimation in conventional FPA systems, due to the highly non-linear relationship between the antenna positions and the wireless channels in practice as well as the unaffordable overhead to perform exhaustive channel estimation at any position. To tackle this challenge, the authors of \cite{ma2023compressed} and \cite{xiao2024channel} proposed a compressed sensing (CS)-based algorithm for MAs by leveraging channel sparsity. The authors of \cite{zhang2023successive} proposed a successive Bayesian reconstructor (S-BAR) without the need of model-driven estimators as in \cite{ma2023compressed} and \cite{xiao2024channel}. The authors of \cite{zhang2024learning} employed the deep-learning tool for channel mapping with an asymmetric graph masked autoencoder (AGMAE)-based architecture. As for MIMO MA systems, a tensor decomposition-based
method was proposed in \cite{zhang2024channel3} for estimating the multi-path channel components. However, most of the above works only aim to reconstruct the real-valued channel gain map, rather than the complex-valued small-scale channel map. Moreover, all of the above works only consider a one-dimensional (1D) or two-dimensional (2D) transmit/receive moving region. For the small-scale channel mapping in a three-dimensional (3D) moving region, it still remains an open and unsolved problem.

\begin{figure}[t]
\centering
\includegraphics[scale=.4]{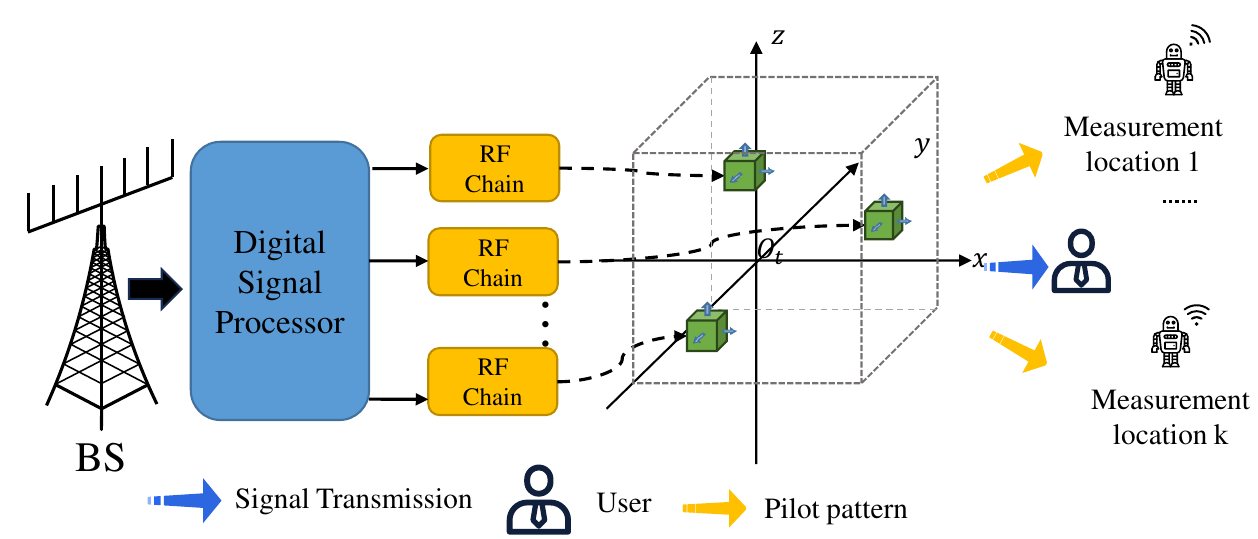}
\caption {MA-assisted MISO communication system.}
\vspace{-12pt}
\label{1}
\vspace{-12pt}
\end{figure}
Motivated by the above, we investigate the complex-valued small-scale channel map estimation problem for a 3D transmit region in this paper. We consider an MA-assisted multiple-input single-output (MISO) communication system, where a base station (BS) is equipped with multiple MAs that can move flexibly within a 3D transmit region, as shown in Fig. \ref{1}. To facilitate the channel map estimation, we first discretize the 3D transmit region into a multitude of sampling points, aiming to acquire an accurate CSI estimation for each sampling point. To avoid an exhaustive channel estimation for all sampling points, we propose a partial channel map estimation protocol with hybrid offline and online stages. In the offline stage, we select a set of MA training positions and receiver locations in the considered environment to estimate their associated CSI. Following this, we train a convolutional neural network (CNN) based on the estimated CSI. Then, the pre-trained CNN is used to
reconstruct the full channel map during real-time transmission based on a finite number of channel measurements taken at several selected sampling points within the 3D movement region. Numerical results based on the 3D field-response channel model demonstrate that our proposed scheme can achieve a higher CSI estimation accuracy than other benchmark schemes.

\begingroup
\allowdisplaybreaks
\section{System Model and Estimation Protocol}
\label{System Model and Protocol}
\subsection{System Model}
As shown in Fig.\,\ref{1}, we consider the downlink of an MA-assisted MISO communication system from a BS to a user. We assume that the BS and the user are equipped with $L$ MAs and a single FPA, respectively. Unlike the existing works assuming a 1D or 2D movement region, we consider that the positions of the $L$ MAs can be flexibly adjusted within a 3D region, denoted as ${\cal C}_t$, which is a cubic space of the size $W\lambda\times W\lambda\times W\lambda$ , where $\lambda$ denotes the wavelength, and $W$ denotes the normalized length per dimension. For convenience, we establish a 3D Cartesian coordinate system, with its origin denoted as $\boldsymbol{o}_t=[0,0,0]^T$, as shown in Fig.\,\ref{1}. Let $h(\boldsymbol{t})$ denote the channel response from an MA to the user when it is deployed at ${\mv t} \in {\mathbb R}^{3 \times 1}$ in ${\cal C}_t$.

To facilitate the channel map estimation, we discretize each dimension of ${\cal C}_t$ into $N$ sampling points, with a sampling spacing of $d=\frac{1}{N-1}W\lambda$. This thus results in $N^{3}$ sampling points in total, which are evenly distributed within ${\cal C}_t$. Let ${\tilde{\boldsymbol{t}}}_{i,j,k} \in {\mathbb R}^{3 \times 1}$ denote the coordinate of the $(i,j,k)$-th sampling point. Consequently, the channel map from the BS to the user can be expressed as $\boldsymbol{\mathcal{H}}\triangleq \{h(\tilde {\boldsymbol{t}}_{i,j,k})|1\le i,j,k \le N\}$, termed {\it{3D small-scale channel map}}.
\vspace{-5pt}
\subsection{Channel Map Estimation Protocol}
Our goal is to reconstruct the 3D channel map $\boldsymbol{\mathcal{H}}$. However, it is noted that the size of the channel map may become prohibitively large as $N$ increases in order to enhance the map resolution. Consequently, conducting exhaustive channel estimation for each sampling point becomes practically unfeasible. Fortunately, the channel coefficients at adjacent sampling points may exhibit spatial correlation due to their similar geometric relationships with environmental scatterers. Inspired by this, we propose a partial channel map estimation protocol with hybrid offline and online stages using a finite number of channel measurements, as shown in Fig.\,\ref{protocol} and described below. 
\begin{figure}[!t]
\centering
\includegraphics[scale=.84]{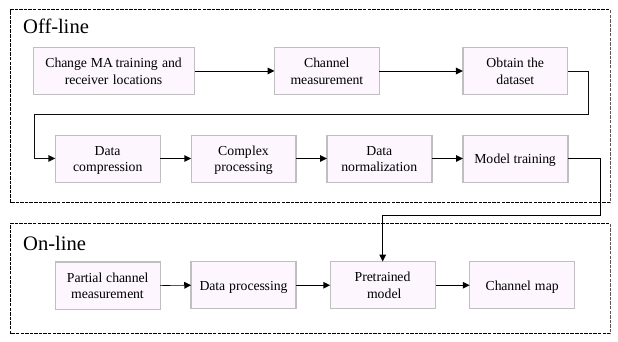}
\caption{Proposed small-scale channel map estimation protocol.}
\vspace{-12pt}
\label{protocol}
\end{figure}

\subsubsection{Offline stage}
As shown in Fig. \ref{1}, we select a set of receiver locations in the considered environment for channel training (e.g., by deploying several mobile robots or fixed sensors). Specifically, the BS adjusts the positions of its MAs and estimates the associated channel coefficients for each selected receiver location in the environment.{\footnote{As there are $L$ transmit MAs, we can measure the channels for $L$ MA training position each time.}} Let $Q$ and $M$ denote the numbers of the MAs' training positions and the selected receiver locations in the environment during the offline channel measurements, respectively. As such, we can obtain $QM$ channel measurements in total, denoted as $h_m(\boldsymbol{t}_q)$, $m=1,2,\cdots,M$ and $q=1,2,\cdots,Q$, where ${\mv t}_q \in {\mathbb R}^{3 \times 1}$ denotes the $q$-th MA training position. These $QM$ measurements are subsequently divided into training and testing sets for CNN-based training. The detailed procedures for data preprocessing and model training will be provided in Section III.

\subsubsection{Real-time stage}
Based on the pretrained model in the offline stage, in the real-time transmission to the user, the BS estimates partial CSI for several MA positions in ${\cal C}_t$, which are then input into the pretrained model to reconstruct the overall channel map $\boldsymbol{\mathcal{H}}$. In this paper, we assume a quasi-static environment, e.g., smart homes and factories, such that the offline training results can be applied in real time over a certain period. In fast-varying scenarios, the offline training results can also be utilized to ensure long-term system performance by capturing the information about the static/dominant paths.

\section{Proposed Channel Map Estimation Algorithm}
\label{CMEA}
In this section, we present the details of our proposed channel map estimation algorithm, including offline channel measurements, data processing, model training, and real-time implementation.

\subsection{Offline Channel Measurements}\label{OCM}
To capture more comprehensive environmental features, we uniformly select a set of sampling points in ${\cal C}_t$ as the MA training positions. In particular, we introduce a spacing parameter $\alpha$ ($\alpha \ge 2$) to adjust the index difference between two adjacent selected sampling points (and thus the total number of the selected points). Hence, there are $Q= \bar{N}^3$ sampling points selected as training positions, with $\bar{N} \triangleq \lceil \frac{N}{\alpha} \rceil^3$, where $\lceil \cdot \rceil$ denotes the ceiling function. It follows that the total number of MA training positions decreases as $\alpha$ increases.
Hence, there exists an inherent complexity-accuracy trade-off in adjusting $\alpha$, since a smaller/larger $\alpha$ is required if a higher/lower accuracy in the 3D channel map is desired. For any given set of selected sampling points, we can estimate their associated point-wise CSI by utilizing pilot-based channel estimation algorithms, e.g., least squares (LS) and minimum mean square error (MMSE) estimators.
\vspace{-6pt}

\subsection{Data Processing}\label{data process}
Next, we process the measured $QM$ point-wise CSI following three steps: complex number processing, data normalization and data compression. For the $m$-th receiver location, let $\mv{H}_m\in\mathbb{C}^{{N}\times{N}\times{N}}$ denote the measured (incomplete) small-scale channel map, where the channel coefficients at the $N^3-\bar N^3$ non-selected sampling points are set as zero. Furthermore, since the neural network can only process real values, the complex CSI is separated into real and imaginary parts, resulting in two 3D CSI matrices each with a dimension of ${N}\times{N}\times{N}$, i.e., $\Re\left\{{\mv H}_m\right\}$ and $\Im\left\{{\mv H}_m\right\}$, $m=1,2,\cdots,M$. In order to enable the network to extract environmental features more accurately, we also compute the estimated channel gain in decibel (dB) in an element-wise manner, resulting in another 3D matrix of the same size, denoted as ${\mv H}_{m,3}=[10\lg\left|h_{m}(i,j,k)\right|^2]_{i,j,k}$, where $h_m(i,j,k)$ denotes the $(i,j,k)$-th entry of ${\mv H}_m$. 

Next, a linear transformation is applied to adjust the data, effectively removing any negative values from the dataset. This operation can compress the data to a manageable size without altering the distribution of the original dataset. Furthermore, it simplifies the calculation of the matrix inverse involved. Mathematically, the linear transformation is expressed as
\begin{equation}
	\label{linear}
    \begin{aligned}
         &\tilde{{\mv H}}_{m,1}={a_1\Re\{{\mv H}_m\}+\delta_1}
         \\
        &\tilde{{\mv H}}_{m,2}={a_2\Im\{{\mv H}_m\}+\delta_2}
        \\
        &\tilde{{\mv H}}_{m,3}={a_3{\mv H}_{m,3}+\delta_3},
    \end{aligned}
\end{equation}
where $a_i$ and $\delta_i, i=1,2,3$ are constants selected based on the characteristics of the data, typically determined by the minimum and maximum values of the dataset. We then concatenate the three 3D matrices in (\ref{linear}) into a four-dimensional (4D) matrix for the $m$-th measurement location, i.e., $\boldsymbol{X}_m\triangleq\left[\tilde{\mv{H}}_{m,1},\tilde{\mv{H}}_{m,2},\tilde{\mv{H}}_{m,3}\right]\in {\mathbb R}^{N \times N \times N \times 3}, m=1,2,\cdots,M$. Finally, the 4D matrix $\boldsymbol{X}_m$ is compressed to reduce the amount of calculation and prevent the gradient from vanishing. Let $f_{\mathrm{cp}}(\cdot)$ denote the compression function. For data compression, the elements in $\boldsymbol{X}_m$ with zero values are removed from $\boldsymbol{X}_m$, which turns $\boldsymbol{X}_m$ into another 4D matrix with a smaller size of $\bar{N}\times\bar{N}\times\bar{N}\times3$. As a result, $f_{\mathrm{cp}}(\cdot)$ can be characterized as a mapping from $\mathbb{R}^{N\times N\times N\times 3}$ to $\mathbb{R}^{\bar{N}\times\bar{N}\times\bar{N}\times3}$.\vspace{-9pt}

\subsection{Model Training}\label{training}
After the above data processing, we can commence training the network. However, in practical scenarios, we cannot obtain the ground-truth channel map, $\boldsymbol{\mathcal{H}}$. Hence, we use part of the measurements as the input subset and another part as the label subset. The input subset is fed into the neural network to predict the channel map, while the label subset is used to evaluate the estimation accuracy and update the network weights. To identify the input and label subsets, we define two binary matrices, i.e., $\boldsymbol{B}^{\mathrm{I}},\boldsymbol{B}^{\mathrm{L}} \in \{0,1\}^{\bar{N} \times \bar{N} \times \bar{N}}$, where $\boldsymbol{B}^{\mathrm{I}}{(n_1,n_2,n_3)} = 1$ (or $\boldsymbol{B}^{\mathrm{L}}{(n_1,n_2,n_3)} = 1$) if the associated
CSI at the $(n_1,n_2,n_3)$-th sampling point is used as the input
(or label) subset, $n_1, n_2, n_3 = 1,\cdots,\bar{N}$; otherwise, it is set to zero. It is evident to see $\boldsymbol{B}^{\mathrm{I}}+\boldsymbol{B}^{\mathrm{L}}={\bf 1}$, where ${\bf 1}$ is an all-one 3D matrix. To align the dimension of $\boldsymbol{B}^{\mathrm{I}}$ or $\boldsymbol{B}^{\mathrm{L}}$ with $\boldsymbol{X}_m$, we concatenate its three identical versions and define a 4D mask matrix $\boldsymbol{S}^i=\left[\boldsymbol{B}^i,\boldsymbol{B}^i,\boldsymbol{B}^i\right]\in \{0,1\}^{N \times N \times N \times 3}, i\in\{\mathrm{I},\mathrm{L}\}$. 
\begin{figure}
	\centering
	\includegraphics[scale=.26]{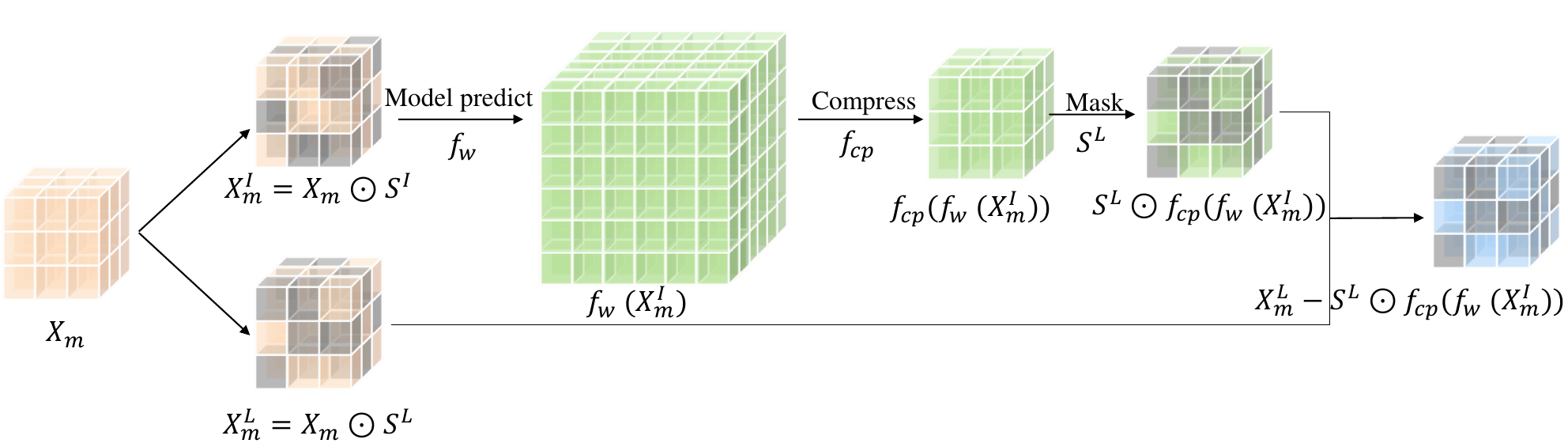}
	\caption{Illustration of the loss function calculation}
	\vspace{-12pt}
	\label{lossfun}
    \vspace{-12pt}
\end{figure}
\begin{figure*}[!t]
\centering
\includegraphics[scale=0.85]{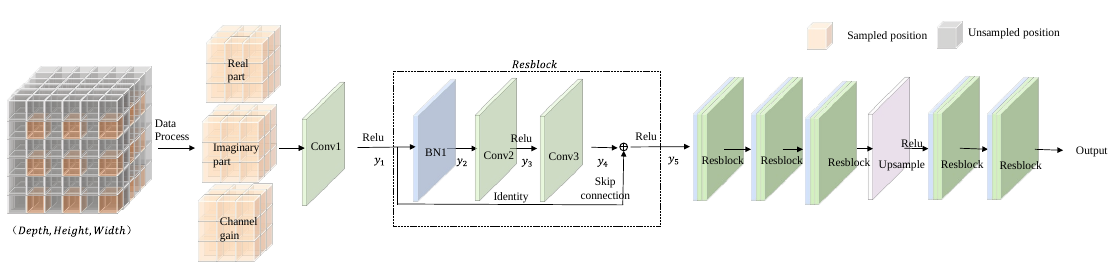}
\caption{Architecture of the proposed CNN}
\vspace{-12pt}
\label{cnn}
\end{figure*}

Then, the input and label data for the $m$-th measurement location can be respectively expressed as
\begin{equation}
\label{Xm}
\boldsymbol{X}_m^{\mathrm{I}}=\boldsymbol{X}_m\odot\boldsymbol{S}^{\mathrm{I}},\quad\boldsymbol{X}_m^{\mathrm{L}}=\boldsymbol{X}_m\odot\boldsymbol{S}^{\mathrm{L}},
\end{equation}
where $\odot$ denotes the Hadamard product, and the mask matrices help to filter the data for network input and labeling, respectively, as illustrated in Fig.\,\ref{lossfun}. Based on (\ref{Xm}), we can define the following mean square error (MSE) as the loss function for training, i.e.,
\begin{equation}
\label{loss}
\mathcal{L}_{\mathrm{MSE}}=\sum_{m=1}^{M}{\left\|\boldsymbol{X}_m^{\mathrm{L}}-\boldsymbol{S}^{\mathrm{L}}\odot f_{\mathrm{cp}}\left(f_W(\boldsymbol{X}_m^{\mathrm{I}})\right)\right\|}_{\mathrm{F}}^2
\end{equation}
where $f_W$ represent the model function. Note that in (\ref{loss}), we only leverage the measurement data in the label set to calculate the MSE, as illustrated in Fig.\,\ref{lossfun}, instead of the ground-truth channel map that is practically unavailable.

Given the loss function in (\ref{loss}), we propose a CNN architecture composed of multiple residual blocks (ResBlock) to estimate the 3D channel map, as shown in Fig.\,\ref{cnn}. Unlike other artificial intelligence (AI)-based methods that rely on global feature interactions or sequential dependencies, CNN inherently exploits the spatial locality of wireless channels through localized convolutional operations. This design aligns with the anisotropic scattering other network characteristics of 3D environments, enabling efficient modeling of adjacent antenna positions influenced by shared scatterers. The weight-sharing mechanism of CNNs further reduces parameter redundancy, ensuring shorter training times compared to architectures with heavy computational graphs or iterative processes. Combined with the residual learning enabled by ResBlocks, which stabilize gradient propagation in deep networks, the proposed CNN can properly balance the trade-off between accuracy and efficiency. 

The ResBlock comprises two convolutional layers, each followed by batch normalization and a ReLU activation function, except the second batch normalization layer. As shown in Fig.\,\ref{cnn}, the input $\boldsymbol{X}_m^{\mathrm{I}} $ is first processed by the initial convolutional layer, with weights \( W_1 \). The output $\boldsymbol{y}_1$ is derived from the input $\boldsymbol{X}_m^{\mathrm{I}} $ after the first complex convolutional layer and the ReLU activation function, i.e.,
\begin{equation}
     \boldsymbol{y}_1 = {\mathrm{ReLU}}(\mathrm{Conv_1}(\boldsymbol{X}_m^{\mathrm{I}}, W_1)),
\end{equation}
where $\mathrm{Conv}_i$ denotes the $i$-th convolutional layer, and the ReLU activation function ${\mathrm{ReLU}}(\cdot)$ is applied to its argument in an element-wise manner, i.e., ${\mathrm{ReLU}}(z) = \max(0, z)$.
Next, $\boldsymbol{y}_1$ is passed through the first batch normalization (BN) layer and the output is given by
\begin{equation}
\boldsymbol{y}_2 = \mathrm{BN_1}(\boldsymbol{y}_1) \triangleq \gamma_1 \left( \frac{\boldsymbol{y}_1 - \mu_{\text{batch}}}{\sqrt{\sigma_{\text{batch}}^2 + \epsilon}} \right) + \beta_1 \ ,
\end{equation}
where $\mu_{\text{batch}}$ and $\sigma_{\text{batch}}^2$ represent the mean and variance of $\boldsymbol{y}_1$, respectively, $\epsilon$ is a sufficiently small amount introduced to prevent division by zero, and $\gamma_1$ and $\beta_1$ are scale and shift parameters, respectively, both of which are to be learned. These parameters are utilized to control the variance and mean of $\boldsymbol{y}_2$.
Then, \( \boldsymbol{y}_2 \) is passed through the second convolutional layer with weights \( W_2 \), i.e.,
\begin{equation}
\boldsymbol{y}_3 ={\mathrm{ReLU}}(\mathrm{Conv_2}(\boldsymbol{y}_2, W_2)).
\end{equation}
Following this, \( \boldsymbol{y}_3 \) is passed through the third convolutional layer with weights \( W_3 \), i.e.,
\begin{equation}\label{y4}
\boldsymbol{y}_4 = \mathrm{Conv_3}(\boldsymbol{y}_3, W_3)),
\end{equation}
Finally, the original input of the Resblock $\boldsymbol{y}_1$ is added to (\ref{y4}) before the ReLU activation function. This addition is a crucial operation in the ResBlock and establishes the residual connection, i.e.,
\begin{equation}
\label{y5}
\boldsymbol{y}_5 = {\mathrm{ReLU}}(\boldsymbol{y}_1 + \boldsymbol{y}_4),
\end{equation}
where \( \boldsymbol{y}_5 \) represents the output of the ResBlock. It is important to note that the operation in (\ref{y5}) helps maintain gradient flow throughout the network and facilitates the training of deeper neural networks. The complete operation of each ResBlock can be summarized as
\begin{equation}
\boldsymbol{y}_5={\mathrm{ReLU}}(\mathrm{Conv_3}(\mathrm{ReLU}(\mathrm{Conv_2}(\mathrm{BN}_1(\boldsymbol{y}_1))))+\boldsymbol{y}_1).
\end{equation}
As shown in Fig. \ref{cnn}, the input will pass through 6 ResBlocks to produce the output.\vspace{-4pt}

\section{3D Channel Model and Simulation Results}
In this section, we simulate and generate the dataset required for training the CNN. Subsequently, we present the simulation results to demonstrate the effectiveness of our proposed 3D channel map estimation protocol.

\subsection{3D Field-Response Channel Model for Data Generation}
\begin{figure}[t]
\centering
\includegraphics[scale=.27]{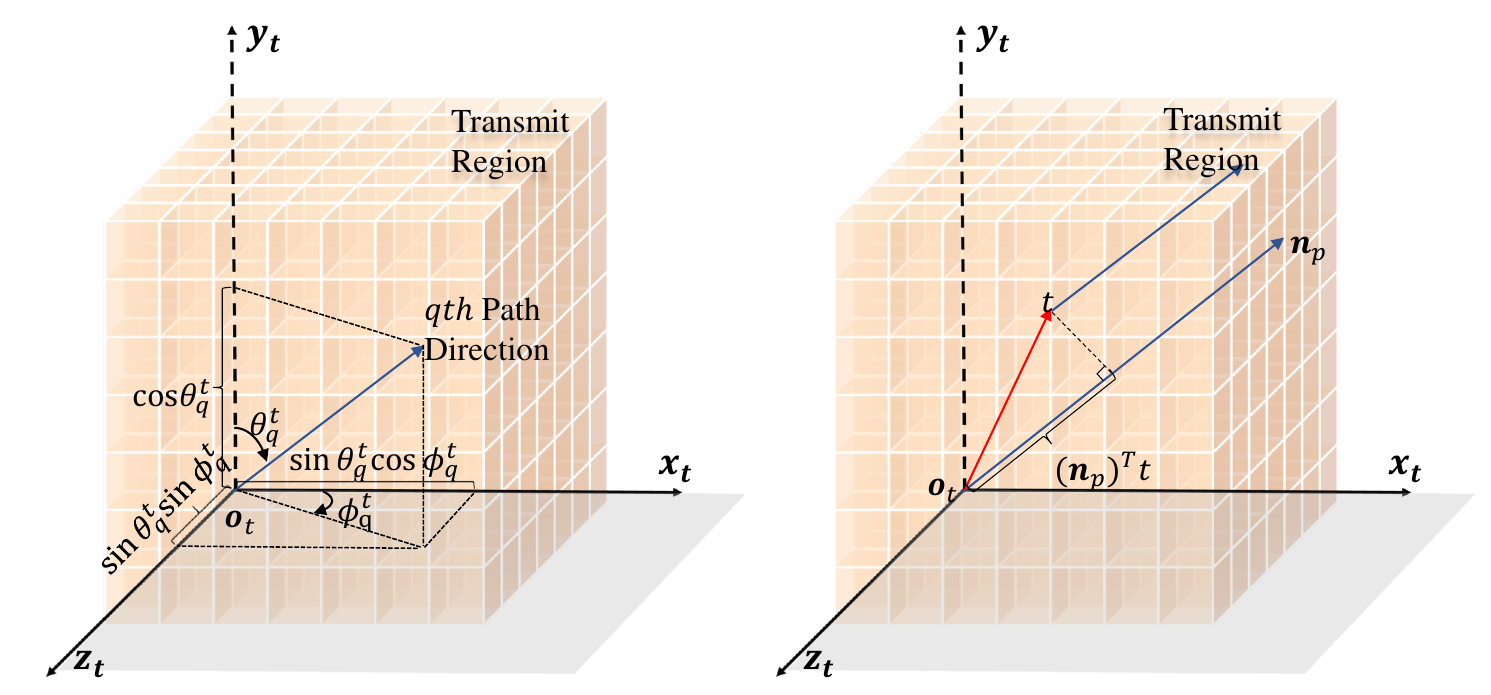}
\caption{3D field-response based channel model}
\vspace{-10pt}
\vspace{-12pt}
\label{3Dfield}
\end{figure}

As shown in Fig. \ref{3Dfield}, we consider the 3D far-field field-response channel model\cite{zhu2023modeling}. Let $L_t$ denote the total number of transmit multi-path components (T-MPCs). The coordinate of the $q$-th ($q=1,2,\cdots,L_t$) T-MPC is represented as $[x_q^\mathrm{T-MPC},y_q^\mathrm{T-MPC},z_q^\mathrm{T-MPC}]$. Therefore, the elevation and azimuth angles of departure (AoD) for the $q$-th T-MPC, i.e., $\theta_q^t$ and $\phi_q^t$, are given by
\begin{align}
    &\theta_q^t\!=\!\arccos\frac{y_q^\mathrm{T-MPC}}{\sqrt{\left(x_q^\mathrm{T-MPC}\right)^2\!+\!\left(y_q^\mathrm{T-MPC}\right)^2\!+\!\left(z_q^\mathrm{T-MPC}\right)^2}},\nonumber\\
    &\phi_q^t\!=\!\arctan\frac{z_q^{\mathrm{T-MPC}}}{x_q^{\mathrm{T-MPC}}}.\label{angle}
\end{align}
Given (\ref{angle}), the directional vector of the $p$-th transmit path can be represented as $\boldsymbol{n}_p=[\sin \theta_t^p\cos \phi_t^p,\cos \theta_t^p,\sin \theta_t^p\sin \phi_t^p]^T$. For any given position ${\mv t}$ in the transmit region, define 
$\rho_t^p(\boldsymbol{t})=\boldsymbol{n}_p^T(\boldsymbol{t}-\boldsymbol{o}_t)$.
Then, the field response vector (FRV) for this position can be represented as $\boldsymbol{g}(\boldsymbol{t})\triangleq\left[e^{j\frac{2\pi}{\lambda}\rho_{t}^{1}(\boldsymbol{t})},e^{j\frac{2\pi}{\lambda}\rho_{t}^{2}(\boldsymbol{t})},\cdots,e^{j\frac{2\pi}{\lambda}\rho_{t}^{L_{t}}(\boldsymbol{t})}\right]^{T}$.
Denote by ${\mv f}\in\mathbb{C}^{L_r\times L_t}$ the path response vector (PRV) over all transmit paths, where its $p$-th entry denotes the complex gain of the $p$-th transmit path. Consequently, the channel from any 3D position ${\mv t}$ to the receiver can be expressed as $\boldsymbol{h}\left( {\mv t}\right) =\mv{f} ^H\mathbf{\mv g}\left( {\mv t} \right)$.
Here, we assume that the PRV follows the circularly symmetric complex Gaussian (CSCG) distribution, i.e., $[\mv {f}]_p\sim\mathcal{CN}\left(0,\frac{\rho d^{-\beta}}{L_t}\right)$, where $\beta$ is the path loss exponent and $d$ is the distance from the BS to the receiver. Based on the above, we can generate the CSI dataset for model training with different ${\mv t}$ and different receiver locations. Nonetheless, we do not assume any prior knowledge about the structure of the field-response channel model in the CNN training. 

\subsection{Simulation Results}
In the simulation, we set the carrier wavelength $\lambda = 0.2$ m and $W = 4$. The path-loss exponent is $\beta = 3$. During the offline stage, receiver locations vary in a spherical space 20-80 m from the BS. A total of 3240 receiver locations are uniformly chosen in this range. The number of sampling points per dimension of the transmit region is $N = 40$. We add random Gaussian noise to the measurement data to account for noise, with a signal-to-noise ratio (SNR) of 30 dB. The training and testing sets are divided randomly, with the testing set accounting for 10$\%$ of the total samples. During the training process, the cosine annealing method is adopted to dynamically adjust the learning rate, with an initial learning rate of 0.001. The Adam optimizer is chosen for the optimization task. To evaluate the practical efficiency of the proposed framework, we analyze the computational overhead of the CNN model.  The training process, conducted on an NVIDIA RTX 4090 GPU, completes in approximately 3.5 hours over 200 epochs (with a batch size of 32) using a dataset with 3240 receiver locations. 

\begin{figure}[!t]
    \centering
    \includegraphics[width=6cm]{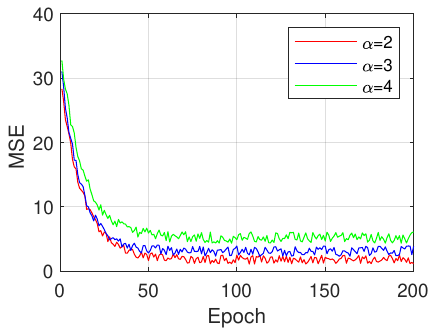}
    \caption{Test losses with different subsampling rates, $\alpha$.}
   \vspace{-12pt}
    \label{testloss}
\end{figure}
First, Fig. \ref{testloss} shows the test error during the model training, i.e., the loss between the predicted data and the label data $\boldsymbol{X}_m^{\mathrm{L}}$ under different values of $\alpha$. It is observed that the test loss gradually decreases with the number of epochs and ultimately converges around 1.815, 3.129, and 5.234 for $\alpha=2, 3$, and $4$, respectively, which correspond to 12.5$\%$, 3.7$\%$, and 1.56$\%$ of the total sampling points, respectively. As expected, a larger value of $\alpha$ gives rise to a higher estimation accuracy, at the cost of the channel training overhead. The above observations demonstrate the effectiveness of the proposed CNN training for small-scale channel map estimation.\vspace{-9pt}

\begin{figure}[h]
	\centering
	\subfloat[Ground-truth channel gain map.]{
		\includegraphics[width=0.155\textwidth]{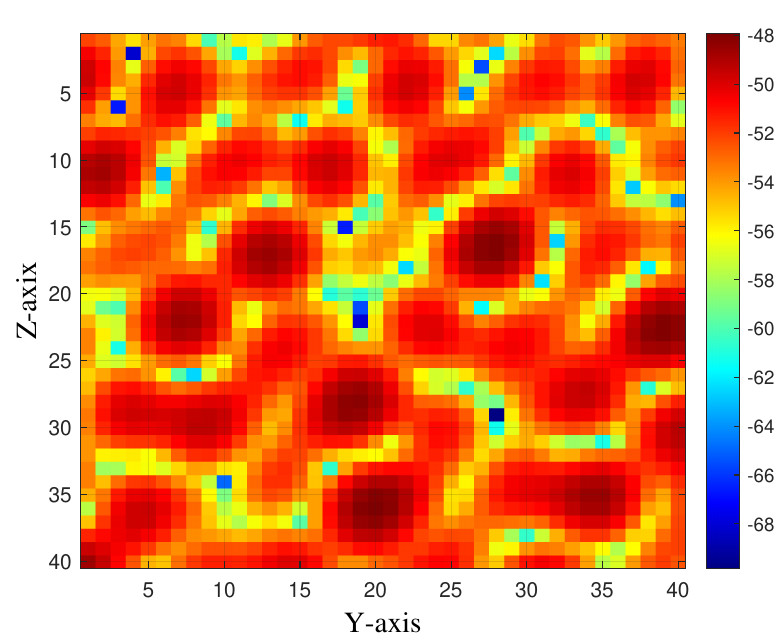}\!
		\label{fig:sub1}
		\includegraphics[width=0.155\textwidth]{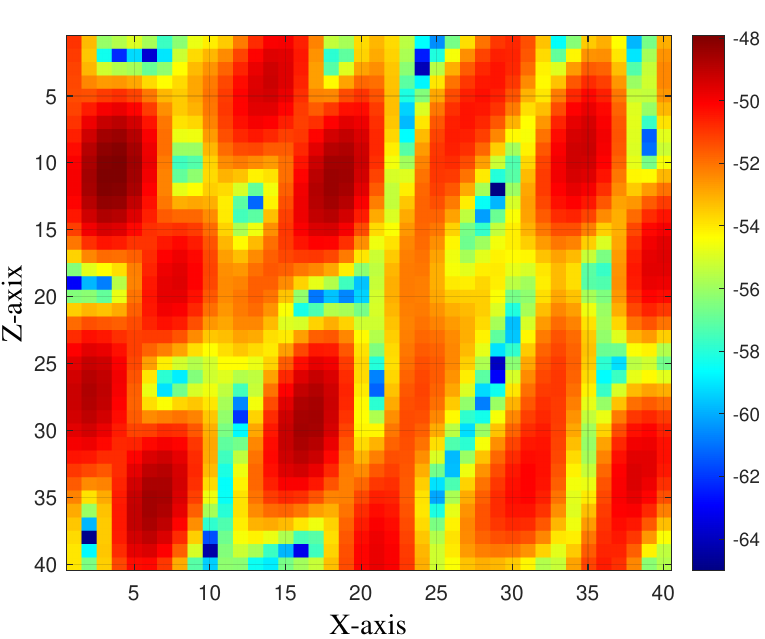}\!
		\label{fig:sub2}
		\includegraphics[width=0.155\textwidth]{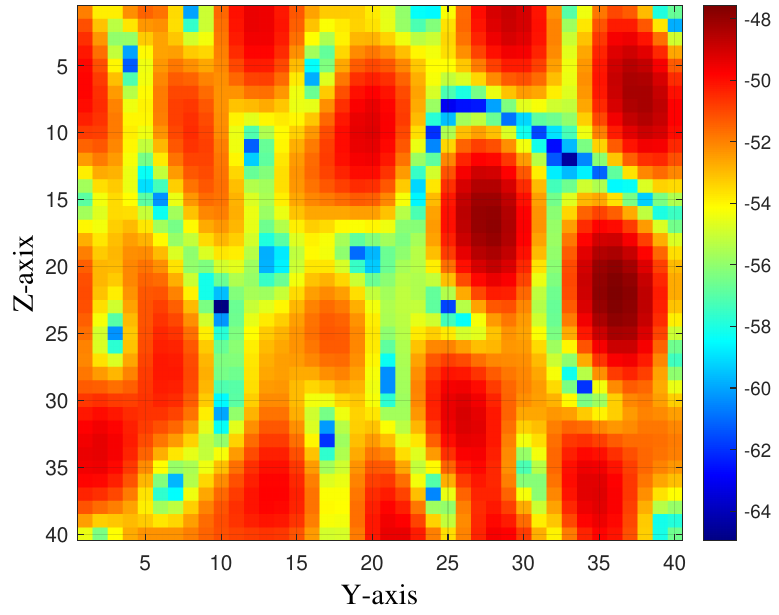}
		\label{fig:sub3}
	}
	\vspace{-9pt}
	\\
	\subfloat[Estimated channel gain map by the proposed scheme.]{
		\includegraphics[width=0.155\textwidth]{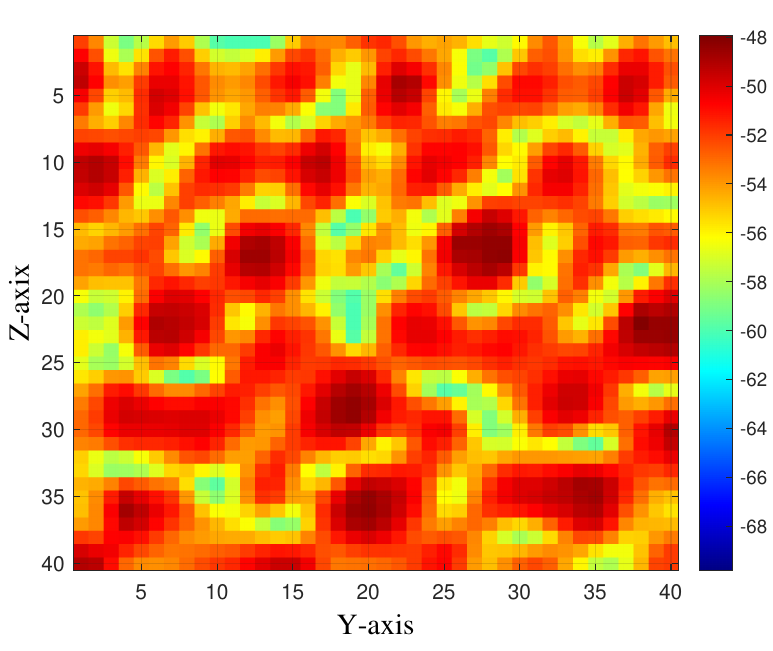}\!
		\label{fig:sub4}
		\includegraphics[width=0.155\textwidth]{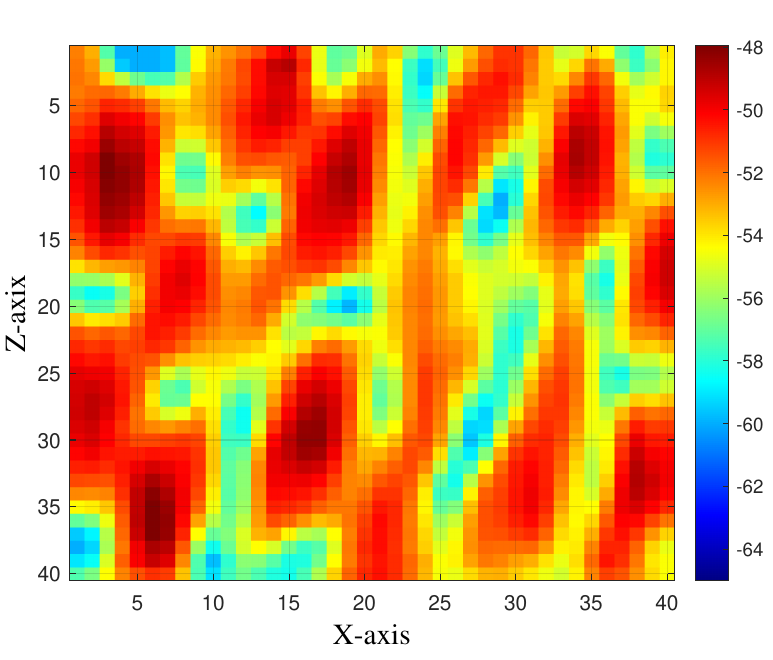}\!
		\label{fig:sub5}
        \includegraphics[width=0.155\textwidth]{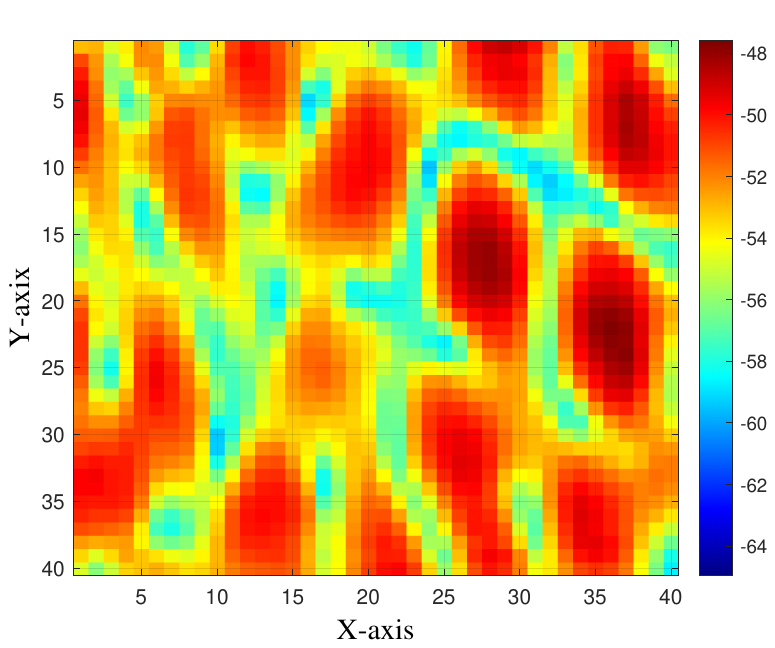}
		\label{fig:sub6}
	}
	\vspace{-12pt}
	\\
	\subfloat[Estimated channel gain map by the trilinear interpolation.]{
		\includegraphics[width=0.155\textwidth]{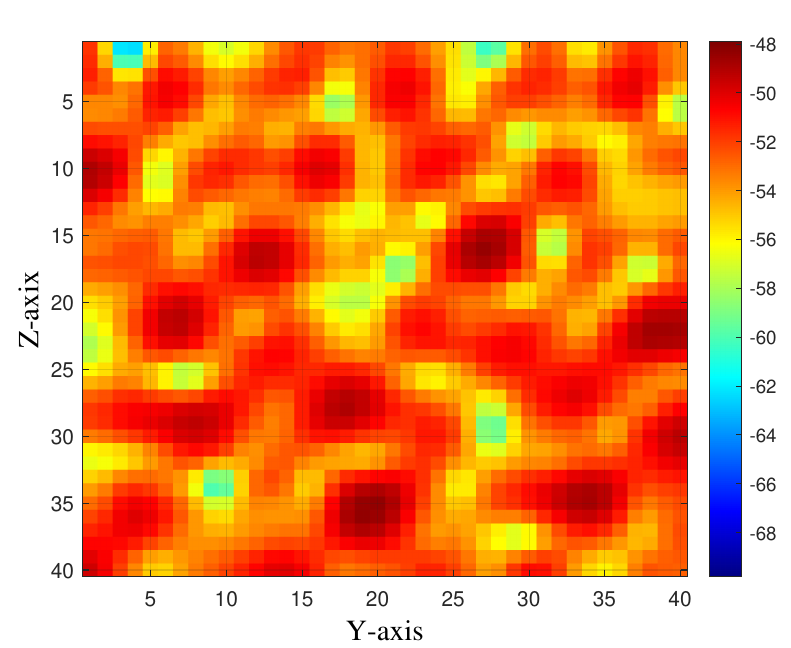}\!
		\label{fig:sub7}
		\includegraphics[width=0.155\textwidth]{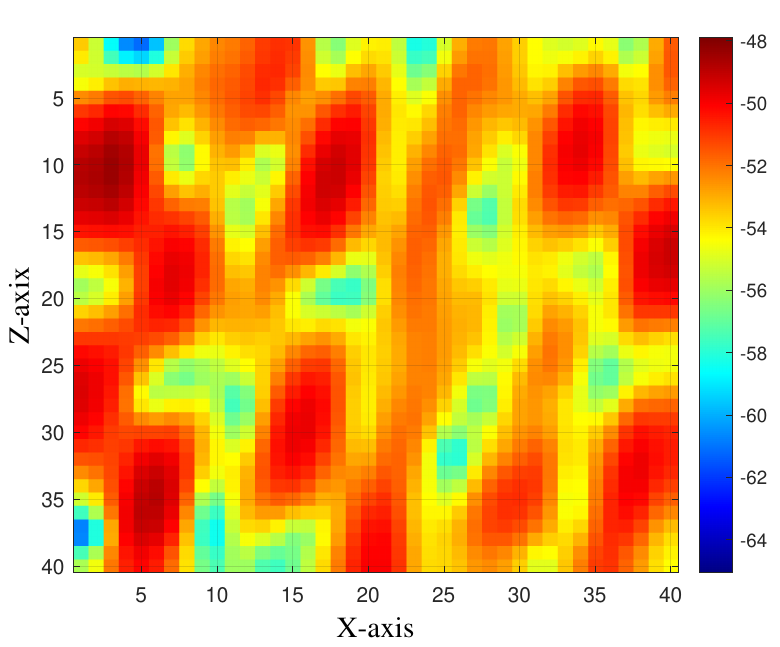}\!
		\label{fig:sub8}
		\includegraphics[width=0.155\textwidth]{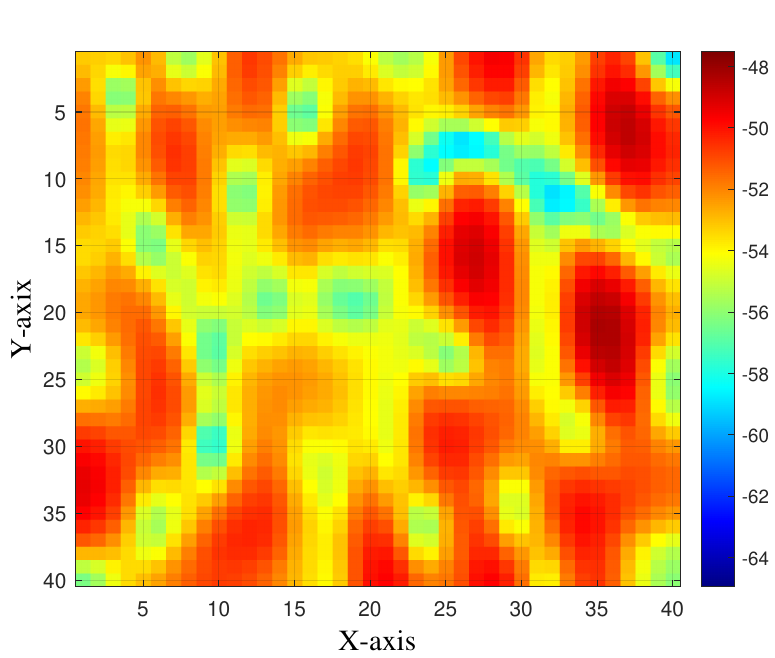}
		\label{fig:sub9}
	}
	\caption{Ground-truth and estimated 2D channel gain maps by different schemes.}
	\vspace{-12pt}
	\label{heat}
\end{figure}
Next, to visualize the efficacy of the proposed scheme, Fig.\,\ref{heat} shows the ground-truth channel gain maps and those by the proposed CNN-based estimation scheme and the benchmark trilinear interpolation method\cite{kummer2010octomag}, which performs 3D linear interpolation based on the measured CSI only. The subsampling rate is set to $\alpha=4$ in the proposed scheme. Due to the 3D nature of the channel gain map, we show its 2D slices at $X=20$, $Y=20$ and $Z=20$, respectively (from left to right). In addition, we show the MSEs achieved by the proposed and the benchmark scheme in Table \ref{tab:performance}. It is observed from Fig.\,\ref{heat} that the proposed CNN-based estimation scheme can achieve a more accurate reconstruction of the details in the ground-truth channel gain map, especially for the first two slices. This superiority is attributed to the adaptive learning capability of 3D convolutional kernels for anisotropic scattering environments in the proposed scheme. Moreover, it is observed from Table \ref{tab:performance} that the proposed scheme can reduce the MSE of the benchmark scheme by 66.9\% (from 17.3984 to 5.7569) for the 3D channel gain map. In addition, an MSE reduction over 60$\%$ can be observed for the sliced 2D channel gain maps as well. The inaccuracy of the trilinear interpolation implies the limitation of its linear operations for nonlinear propagation environments.\vspace{-6pt}
\newcolumntype{C}[1]{>{\centering\arraybackslash}p{#1}}
\def\colwidth{\dimexpr0.3333\columnwidth-2\tabcolsep-1.3333\arrayrulewidth\relax}
\begin{table}[!t]
	\centering
	\caption{Comparison of MSEs by different schemes}
	\label{tab:performance}
	\begin{tabular*}{\columnwidth}{@{}C{\colwidth}C{\colwidth}C{\colwidth}@{}} 
		\toprule
		\textbf{Dimension} & \textbf{CNN} & \textbf{Trilinear} \\
		\midrule
		2D, $X=20$  & 5.7953  & 16.0792 \\
		2D, $Y=20$  & 4.9643  & 18.2365 \\
		2D, $Z=20$  & 7.9457  & 19.7637 \\
		\textbf{3D} & \textbf{5.7569} & \textbf{17.3984} \\
		\bottomrule
	\end{tabular*}
\end{table}\vspace{-24pt}

\section{Conclusion}
\label{conclusion}
In this paper, we investigated the 3D small-scale channel map estimation problem in an MA system. To achieve an accurate channel map reconstruction without the need of exhaustive channel estimation, we proposed a partial channel estimation protocol with hybrid offline and online processing. A CNN-based architecture was introduced to recover the overall channel map using the partially estimated channel coefficients for a set of MA training positions and receiver locations. Numerical results based on the 3D field-response channel model demonstrated the efficacy of our proposed scheme in estimating the complex-valued channel map and the superiority to the benchmark scheme.\vspace{-6pt}

\bibliographystyle{IEEEtran}
\bibliography{reference} 
\end{document}